\documentstyle[prb,aps,preprint]{revtex}

\begin{document}

\title{First order phase transition of the vortex lattice in twinned 
YBa$_2$Cu$_3$O$_7$ single crystals in tilted magnetic fields }
\author{B.Maiorov, G.Nieva, and E.Osquiguil}
\address{Comisi\'{o}n Nacional de Energ\'{\i}a At\'{o}mica-Centro 
At\'{o}mico.
Bariloche and Instituto Balseiro, 8400 Bariloche, Argentina.}

\date{August 23, 1999}
\maketitle

\begin{abstract}
We present an exhaustive analysis of transport measurements performed in {\it 
twinned} YBa$_2$Cu$_3$O$_7$ single crystals which stablishes that the vortex 
solid-liquid transition is first order when the magnetic field $H$ is applied at 
an angle $\theta$ away from the direction of the twin planes. 
We show that the resistive transitions are {\it hysteretic} and the V-I curves 
are 
non-linear, displaying a characteristic s-shape at the melting line $H_m(T)$, 
which scales as $\varepsilon(\theta)H_m(T,\theta)$. These features are gradually 
lost when the critical point $H^{*}(\theta)$ is approached. Above 
$H^{*}(\theta)$ the V-I characteristics show a linear response in the 
experimentally accessible V-I window, and  the transition 
becomes reversible. Finally we show that the first order phase transition takes 
place 
between a highly correlated vortex liquid in the field direction and a solid 
state of unknown symmetry. As a consequence, the available data 
support the scenario for a {\it vortex-line melting} rather than a {\it vortex 
sublimation} as recently suggested [T.Sasagawa et al. PRL 80, 4297 (1998)].
\end{abstract}

\pacs{PACS 74.60.Ge;74.60.Ec; 74.62.Dh; }

\section{Introduction}

During the last years, vortex physics in high-temperature superconductors has become
a major topic of research. The main reason for this interest is that the 
interplay between thermal fluctuations, anisotropy and disorder determine the 
existence of several vortex phases in the magnetic phase diagram of these 
materials. These phases are separated by different kinds of thermodynamic 
transitions. In particular, it 
is now well established that in {\it clean} YBa$_2$Cu$_3$O$_7$ crystals the 
vortex solid transforms into a liquid through a first order phase transition 
\cite{safar,welp,schilling}. This thermodynamic transition shows up in the 
transport properties as a sharp drop or ``kink" in the resistivity, $\rho (T)$, 
at 
the melting temperature $T_m (H,\theta)$ \cite{safar,kwok}, and has an
hysteretic behaviour both in temperature and field \cite{safar} which 
corresponds to a superheating of the solid phase \cite{melissa}. 

When correlated disorder is present in the sample, as for example in twinned 
YBa$_2$Cu$_3$O$_7$ crystals, the transition transforms into second order 
\cite{kwok} and occurs between a solid called Bose-Glass \cite{nelson,grigera} 
and an entangled vortex liquid. The Bose-Glass phase, existing only within a 
small angular region around the direction of the correlated defects ($\theta = 
0$), is characterized by a universal behaviour of the non-linear and linear 
resistivity with well defined critical exponents \cite{grigera}. At the Bose-
Glass transition temperature, $T_{BG} (H,\theta)$, the presence of correlated 
defects introduces changes in the thermodynamic properties of the mixed state 
that go beyond the mass anisotropy approximation. In contrast to $T_m 
(H,\theta)$ which smoothly follows the angular dependence given by the 
anisotropy, $T_{BG} (H,\theta)$ shows a sharp cusp \cite{nelson,grigera}around 
$\theta = 0$. Within this small angular region, $\rho (T)$ has a smooth 
temperature dependence near the transition, while at larger angles a similar 
kink to that observed in untwinned crystals develops \cite{kwok,grigera,morre}. 
Using the Lindeman criterion and the scaling rules for anisotropic 
superconductors \cite{blatter} Kwok and co-workers \cite{kwok} fitted the 
angular 
dependence of the temperature at which the kink occurred. This fit was 
interpreted as an indication of a recovery of the melting transition in the 
vortex system when the magnetic field was tilted away from the planar defect 
direction \cite{kwok}.

Using this interpretation, Langan et al. showed in a recent paper \cite{langan} 
that the kink in $\rho(T)$ is suppressed for fields larger than a certain 
field $H^{*} (\theta)$. This behaviour is similar to that observed in untwinned 
crystals where a critical end point for the melting line has been identified 
\cite{safar3}, and gives support to the occurrence of a first order solid-
liquid 
transition when the field is rotated away from the twin planes.

An interesting finding in oriented twinned crystals in inclined magnetic fields 
was done by Morr\'e et al. \cite{morre}, who showed that the transport 
properties of the liquid state at angles well beyond the so-called depinning 
angle, $\theta _d$, are quite different from those observed in clean samples. 
The depinning angle concept is commonly used in the literature to indicate the 
angle beyond which the twin boundaries becomes ineffective as a correlating 
potential.
Morr\'e et al. showed, however, that the twin boundary potentials continue to 
induce vortex velocity correlation even for $\theta >> \theta _d$. Their results 
demonstrated that the vortex liquid remains correlated in the field direction 
above the resistivity kink temperature. This behaviour is in sharp contrast to 
that observed in untwinned samples where the vortex velocity correlation in the 
field direction is lost at the melting temperature \cite{lopez}. The important 
difference between the characteristics of the two liquids was used to cast 
doubts about the interpretation of the resistivity kink in twinned crystals as a 
manifestation of a first order transition \cite{morre}.

In order to investigate further the nature of the vortex solid-liquid transition 
in the presence of correlated defects at finite angles, we studied the vortex 
dynamics in YBa$_2$Cu$_3$O$_7$ crystals with oriented twin boundaries. We performed 
exhaustive transport measurements as a function of magnetic field, $H$, and 
angle, $\theta$, with respect to the twin boundary direction. 
Characterizing the resistivity kink near the transition by its width, we show 
that it remains 
almost constant as $H$ is increased until a field $H^{*}(\theta)$ is reached. 
Below this field the V-I characteristics are non-linear and, what is more 
important, {\it the resistive transitions are hysteretic}. Moreover, our results 
indicate that the hysteresis corresponds to a superheating of the vortex 
lattice, in agreement with the results obtained by Charalambous et al. \cite 
{melissa} Above $H^{*}$ which scales as $\varepsilon(\theta)H^*(\theta)$, 
the kink width suddenly increases, the V-I characteristics become linear and 
{\it the hysteresis disappears}.
We find that the resistivity kink height follows a universal 
behaviour when the reduced variable $\varepsilon(\theta) H$ is used, in 
contrast to the results of Langan et al.\cite{langan}. 
Our data together with those previously reported \cite{kwok,langan}, lead 
us to conclude that when $H$ is tilted away from the direction of the twin planes 
the transition to the vortex liquid state, characterized by a steep jump in the 
resistivity, is indeed a first order phase transition. This transition occurs 
between a highly correlated vortex liquid \cite{morre} and a solid of unknown 
symmetry and therefore supports an interpretarion in therms of line-like melting 
\cite{koshelev} rather than a sublimation of the vortex lattice \cite{sasagawa}. 

\section{Experimental}

We carried out transport measurements on YBa$_2$Cu$_3$O$_7$ single crystals with 
only one family of oriented twins. The crystals were grown using the
self flux method \cite{delacruz-94}. The fully oxygenated samples had typical 
dimensions $1 \times 0.3 \times 0.01 mm^3$, critical transition temperatures 
$T_c = 93.2K$, and transition widths $\Delta T_c \leq 0.6K$. Four parallel 
contacts separated by 150$\mu m$ were 
made with silver epoxy over evaporated gold pads, resulting in contact 
resistances lower than 1$\Omega$. The crystal was mounted onto a 
rotatable sample holder with an angular resolution better than $0.05^\circ$ 
inside a commercial gas flow cryostat with an 18T magnet. The current was 
injected at $45^\circ$ off the twin planes. Transport measurements were 
performed using conventional DC techniques. All resistivity measurements were 
made within the linear regime using a current density $J\leq 3 A/cm^2$.

\section{Results and Discussion}

In Figure \ref{tivstita6t} we show the angular dependence of the temperature at 
which the resistivity of the crystal becomes zero within experimental resolution 
\cite{aclaracion}, at an applied magnetic field of 6T. Similar results were 
obtained for other applied fields. As clearly seen, there is a cusp at small 
angles which is indicative of the Bose-Glass phase, as has been recently 
demonstrated by Grigera et al. \cite{grigera}. In this paper we will concentrate 
on the investigation of the nature of the solid-liquid transition in the angular 
region away from that in which the Bose-Glass phase exists.

In Figure \ref{rts}  we present $R(T)$ data for three different angles, as a 
function of the applied magnetic field. As can be seen, the transitions at low 
fields display a sharp drop to zero which we identify with the characteristic 
resistivity kink. Note that, for a fixed angle, the kink is washed out as the 
field is increased, an effect similar to that observed in ref. [10] and 
to that reported for untwinned crystals with the magnetic field applied parallel 
to the $c$-axis \cite{kwok}. In the case under study, the field value at which 
the transitions starts to broaden depends on the angle between $H$ and the 
correlated defects: the larger the angle, the larger the field at which the kink 
is washed out. This feature is more clearly seen in Figure \ref{anchos} where, 
in order to 
quantify the above mentioned behaviour, we plotted the full width at half 
maximum 
(FWHM) of the temperature derivative, $d\rho/dT$, of the curves shown in Fig. 
\ref{rts} as a function 
of the applied field $H$. It can be seen that for $\theta = 23^\circ$ the 
transitions start to broaden at approximately $H \approx 12 T$, while this 
field value is increased up to 14T for $\theta = 38^\circ$, and to almost 18T 
for $\theta = 58^\circ$. The values of the magnetic field at which the 
transitions start to broaden are independent of the criterion used for the 
definition of the transition width. 

According to the scaling theory of Blatter et al. \cite{blatter}, in an 
anisotropic superconductor a physical magnitude which is a function of angle and 
field should scale as the product $H \varepsilon(\theta)$, where the 
anisotropy factor $\varepsilon (\theta) = \sqrt {cos^2 (\theta) + \gamma^{-2} 
sin^2 (\theta)}$, with $\gamma^2 = m_c/m_{ab}$. Such scaling for the FWHM, with 
$\gamma = 7$, is shown in Figure \ref{anchosesc}, where we have also included 
data for other 
measured angles. Note that all curves collapse onto one and 
that the sudden increase of the transition widths occurs at the same reduced 
field $ H^{*}= H \varepsilon(\theta) \simeq 11T$ for all angles. The sudden 
change in the transition width is indicative of a corresponding change in the 
vortex dynamics below and above the characteristic field $H^{*}$. 

As is widely accepted, one of the most powerful tools to investigate vortex 
dynamics is the measurement of V-I characteristics. We performed such 
measurements in a temperature interval around the transitions for 
magnetic fields below and above $H^{*}$. Typical results for the resistance as a 
function of the applied current at an angle $\theta = 23 ^\circ$ are shown in 
Figure \ref{vis}. In panel (a) we plot the measurements for a field of 8T, lower 
than the 
critical field $H^{*} (23 ^\circ)$. At low currents and high temperatures the 
vortex response is ohmic, changing to a non-linear behaviour as the temperature 
is reduced towards $T_m = 80.4$. Note that in the intermediate temperature 
region, the R(I) curves have a characteristic {\it s}-shape. This feature, which 
has been reported in clean samples where a first order phase transition 
in the vortex lattice was identified \cite{nose2}, is characteristic of all the 
magnetic field region below $H^{*} (23^\circ)$. It is more pronounced as $H$ is 
reduced, becoming less evident as $H$ approaches $H^{*}$. Well above $H^{*}$ the 
$R(T)$ curves are linear up to the maximum current that can be used without producing heating effects. In panel (b) we show the results for $H \simeq H^{*} 
(23 ^\circ)$. Clearly, the vortex response is remarkably different from 
that shown in panel (a). In a wide range of applied currents the response is 
linear. Non-linearities developed at high currents and low 
temperatures and are due to vortex loop excitations \cite{esteban}. One may 
wonder if these non-linearities are a sign of a glass transition taking place at 
lower temperatures, a question that has also been raised concerning the nature 
of the transition above the critical point in clean crystals. If this were the 
case, the tails of the resistive transitions above the critical field should 
follow the scaling behaviour predicted by the theory \cite{fisher}, $R \sim (T-
T_g)^{\nu (z-1)}$. The analysis of our resistivity data in terms of this scaling 
yielded negative results. We cannot discard, however, the occurrence of the 
glass 
transition, since this negative results might be related to the voltage 
resolution we have in our experiments. Safar et al. \cite{safar2} experimentally 
observed the above mentioned scaling in Bi$_2$Sr$_2$CaCu$_2$O$_{8+\delta}$ 
crystals by using SQUID 
picovoltimetry.

The extensive experimental results shown above display special features pointing 
towards
the occurrence of a first order solid to liquid transition in the vortex lattice 
when $H <
H^{*}$. In order to consolidate this scenario, we have performed measurements to 
look for
hysteretic behaviour in the resistive transitions\cite{safar}. Due to 
experimental
constrains, instead of searching for hysteresis in temperature, we performed 
such
measurements at a fixed temperature (regulation better than 5 mK) and sweeping the magnetic field up and down. The results for an angle $\theta = 23 ^\circ$ and a 
measuring current
$I=50 \mu A$ are plotted in Figure \ref{histb}. Panel (a) shows the data for $T 
\simeq
T^{*}$ where $ T^{*}$ is the temperature corresponding to the critical field at 
$\theta =
23 ^\circ$. The arrows indicate the sense in which the field was sweeped. Within
experimental resolution no hysteresis is seen in this region. However, when the 
temperature
is increased in such a way that $T > T^{*}$ a clear hysteretic behaviour in the 
resistive
transition develops, as can be seen in panel (b). It is important to mention 
that the width of the
hysteresis is current independent between $50$ and $100 \mu A$. When the 
applied
current is increased above this value curve A shifts towards curve B and, at 
high enough
currents (above $150 \mu A$), the hysteresis is washed out. Following 
Charalambous et al.
\cite{melissa} we interpret this behaviour as indicative of a superheating of 
the vortex
lattice.

In Figure \ref{htdiag} we compare the superconductor phase diagram for our 
sample with that
obtained in a clean untwinned crystal \cite{safar}. The inset show the raw H-T 
data, for
different angles and the data at zero angle for the untwinned crystal. In order 
to take
into account the anisotropy change as the angle between $H$ and the defects is 
changed, in
the main panel we have used the corresponding scaling field $H 
\varepsilon(\theta)$, while
the reduced temperature scale $t=T/T_c (H=0)$ is used to account for the 
different critical
temperature at zero field of both samples. The collapse of all data on one 
universal curve
not only provides an impressive graphical view of our interpretation of the 
resistivity
kink temperature in twinned samples as a melting temperature, but indicates that 
{\it the
presence of the twin boundary potentials does not modify this temperature}. 
Later on we
will come back to this point, continuing now with the anisotropy dependence of 
other
physical quantities.

In clean crystals a relevant quantity related to the melting transition is the 
kink height at the melting temperature. It has been found to be angle and field 
independent \cite{kwok}, and this particular behaviour has attracted 
theoretical interest \cite{balseiro}. In contrast, we have found that in $45^ 
\circ$ oriented twinned crystals, the kink height is angle and field dependent, 
an observation already reported in ref. \onlinecite{langan}. Since this quantity 
related to the occurrence of a first order phase transition which follows the 
anisotropy, we expect the kink height also to scale with it. In Fig. \ref{vkink} 
we show such scaling. We have plotted the kink height measured at fixed angles 
and increasing the magnetic field (open symbols), together with measurements of 
the 
same quantity but at a fixed applied field (6T) and reducing the angle 
$\theta$(full symbols). 
Clearly all data collapse onto a universal curve. 
The behaviour reported by Langan et al.\cite{langan} is distinctly different.
The reasons for the lack of scaling in their data is related to the fact that 
their measurements at $\theta = 5^\circ$ were taken in an angular region were 
the dissipation at the transition is greatly reduced due to the effect of the 
twin planes on the vortex dynamics. This reduction is easily seen in $V(\theta)$ 
measurements where a dip in the dissipation occurs \cite{morre,langan,fleshler} 
for $\theta \leq 10^\circ$, an angle usually identified with $\theta _d$. One 
may wonder if this change in the vortex dynamics is a consequence of a 
corresponding change in the thermodynamic nature of the transition for $\theta < 
\theta _d$. This is certainly true when the Bose-Glass phase is reached since it 
has been demonstrated \cite{grigera} that in this case the transition is second 
order. However, this phase exists \cite {grigera} below $\theta _{BG} \approx 2^\circ$,
leaving a rather wide angular region ($\theta _{BG} \leq \theta \leq \theta _d$) where the full kink height develops (see e.g. ref. \onlinecite{morre}). One 
interesting possibility \cite{nelson} is that this angular region comprises a 
reentrant liquid phase separating the Bose-Glass from the vortex solid, 
in a similar way to what has been predicted to occur near $H_{c1}$ in clean samples in
\cite{nelson2}. In this case the resistive transitions below the developing kink 
should be linear due to an increased viscosity of the reentrant liquid phase as 
the temperature is lowered. Within this picture $\theta _d$ would be the angle 
below which the reentrant phase starts to develope. Another possibility is that 
the angular region between $\theta _{BG}$ and $\theta _d$) is governed by 
critical fluctuations of the Bose-Glass phase associated to the new 
thermodynamic variable $H_\perp= H sin(\theta)$. If this were the case the so-called 
depinning angle should be interpreted as the critical angle above which the 
first order solid-liquid transition sets in. Although a detailed analysis of the 
nature of the solid-liquid transition in this angular region is out of the scope 
of this paper, further investigation is underway to elucidate this interesting 
issue.

In the following, we would like to comment on a few important points concerning 
the $H-T$ phase diagram shown in Fig. \ref{htdiag}. The first one is related to 
the value of the critical field $H^{*}$ compared to that of the untwinned 
sample. It is well known that in untwinned crystals this field has not a 
universal value. Its magnitude depends on 
the amount of point like disorder present in each crystal \cite{desorden}, the 
larger the disorder the lower the critical field. The results in Figure 
\ref{htdiag} which show $H^{*} \simeq H^{*}_S$ may therefore 
suggest that in tilted magnetic fields the presence of twin boundaries does not 
increase the amount of point like disorder. This suggestion seems to be 
corroborated by our measurements in another oriented twinned crystals and by the 
results shown in ref. \onlinecite{langan} with a value of $H^{*} \simeq 11T$, similar 
to that obtained in our crystals.

The second remark is more fundamental since it is related to the nature of the 
solid and liquid vortex phases in the twinned crystal. In a recent paper 
Sasagawa et al. \cite{sasagawa} suggested that the first order solid-liquid 
phase transition in the vortex system of high temperature superconductors, 
including the less anisotropic YBa$_2$Cu$_3$O$_7$, corresponds to a {\it vortex
sublimation} rather than a {\it line melting}.
Within this scenario the vortex velocity correlation length in the field 
direction should vanish at the melting temperature because the vortex liquid 
phase is formed by uncorrelated pancakes. As already mentioned in the 
introduction, Morr\'e et al. have shown \cite{morre}, using the flux 
transformer contact configuration, that the vortex liquid in tilted magnetic 
fields mantains the vortex velocity correlation in the field direction even at 
rather large angles off the twin planes. Since we have concluded that the 
transition in tilted fields is indeed first order, their results imply that the 
solid transforms into a vortex liquid of {\it correlated lines}, contrary to 
what happens in clean samples. This has a two important implications: first 
the sublimation scenario proposed by Sasagawa et al. \cite{sasagawa} does not 
hold for the melting transition in twinned crystals for tilted magnetic fields, 
and second, although the twin boundary potentials do not affect the melting 
temperature they play an important role in building up the vortex velocity 
correlation in the field direction. This last point indicates that the degree of 
correlation in the vortex liquid does not determine the nature of the 
thermodynamic transition. Therefore one may conclude that the liquid phase in 
clean and untwinned samples (in tilted magnetic fields) is essentially the same 
but with different dynamics due to the different vortex velocity correlation. On 
the other hand, it could also be possible that the nature of the liquid 
phase differs from that in untwinned samples. We speculate that for 
twinned crystals in tilted magnetic fields, the entangled vortex liquid might be 
formed by stair-like correlated lines which are stabilized by the twin boundary 
potential. Within this picture, the solid might have a complicated ordered 
structure also formed by stair-like vortices.

\section{Conclusion}

We have shown that the vortex solid-liquid transition in twinned 
YBa$_2$Cu$_3$O$_7$ crystals for magnetic fields tilted away from the planar 
defect's direction is first order. This thermodynamic transition is reflected in 
the transport properties as a sharp kink in the $R(T)$ curves at the melting 
temperature $T_m (H,\theta)$. The resistive transitions are hysteretic below a 
critical field that scales with the anisotropy of the material, with the 
hysteresis indicating a superheating of the solid phase. The correlated vortex 
liquid phase near the melting temperature could be probably formed by stair like 
entangled lines that maintain their vortex velocity correlation in the field 
direction due to the effect of the twin boundary potential. The structure of the 
vortex solid is unknown.

\section{Acknowledgments}

We acknowledge stimulating discussions and comments from F. De la Cruz. We thank
J.Luzuriaga for a critical reading of the manuscript. This work is partially supported by
ANPCyT, Argentina, PICT 97 No.03-00061-01116, and CONICET PIP 4207. E.O. and G.N
acknowledge financial support from Consejo Nacional de Investigaciones Cient\'{\i}ficas y
T\'ecnicas. 

\bibliographystyle{prsty}

\begin{thebibliography}{10}
\bibitem{aclaracion}
These measurements were performed at low enough currents to inssure that the 
transition width did not change due to non-linearities. This is specially 
important when measuring $\rho (T)$ at large angles.
\end{thebibliography}

\noindent
\begin{figure}
\caption[]{Zero resistance temperature as a function of the angle $\theta$ 
between the applied magnetic field ($H = 6T$) and the twin boundary direction} 
\label{tivstita6t}
\end{figure}

\noindent
\begin{figure}
\caption[]{Resistivity as a function of temperature for different applied 
fields. (a) $\theta = 23 ^\circ $, (b) $\theta = 38 ^\circ $, (c) $\theta = 58 
^\circ$. The curves shown correspond to magnetics fields of H = 0, 2, 3, 5, 7,8, 
9, 10, 11, 12, 13, 14, 15, 16 and 18T.}
\label{rts}
\end{figure}

\noindent
\begin{figure}
\caption[]{Full Width at Half Maximum of the derivatives of the resistive 
transitions as a function of the magnetic field.}
\label{anchos}
\end{figure}

\noindent
\begin{figure}
\caption[]{Full Width at Half Maximum of the derivatives of the resistive 
transitions as a function of the reduced magnetic field $H \varepsilon 
(\theta)$.}
\label{anchosesc}
\end{figure}

\noindent
\begin{figure}
\caption[]{V-I characteristics at an angle $\theta = 23 ^\circ$ for: (a) $H=8$T, 
below the critical field, (b) $H=11$T, just above the critical field. The V-I 
curves were taken at steps of 0.2 K.}
\label{vis}
\end{figure}

\noindent
\begin{figure}
\caption[]{Resistive transitions as a function of $H$ at an angle $\theta = 23 
^\circ$. Curves A were measured sweeping the field up, while curves B were 
recorded by sweeping the field down. (a) $T= 78.52$K, near the critical 
temperature, (b) $T=81.82$K, above the critical temperature.}
\label{histb}
\end{figure}

\noindent
\begin{figure}
\caption[]{Comparison of the melting lines for the twinned crystal in tilted magnetic fields (open symbols) and untwinned crystal with H parallel to c axis after ref. \onlinecite{safar3}. The main panel shows the scaled data, while in the inset in the inset the raw data are plotted.}
\label{htdiag}
\end{figure}

\noindent
\begin{figure}
\caption[]{Normalized height of the resistive kink at the melting temperature as a function of the reduced field. Open symbols correspond to measurement at fixed angles and varying the magnetic field. Solid symbols were measured at 6T varying the angle between $\theta = 20^\circ$ and $\theta = 85^\circ$.}
\label{vkink}
\end{figure}

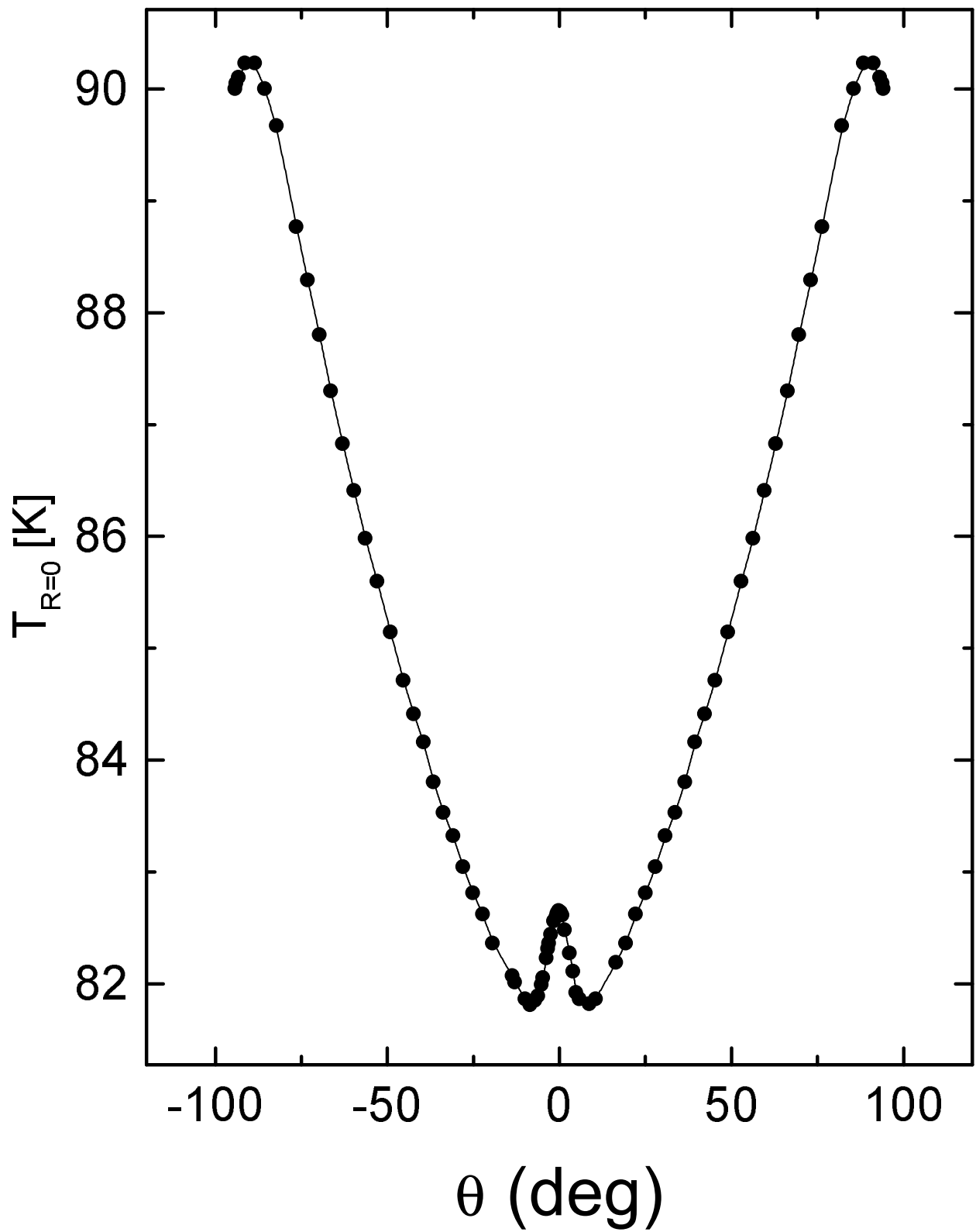
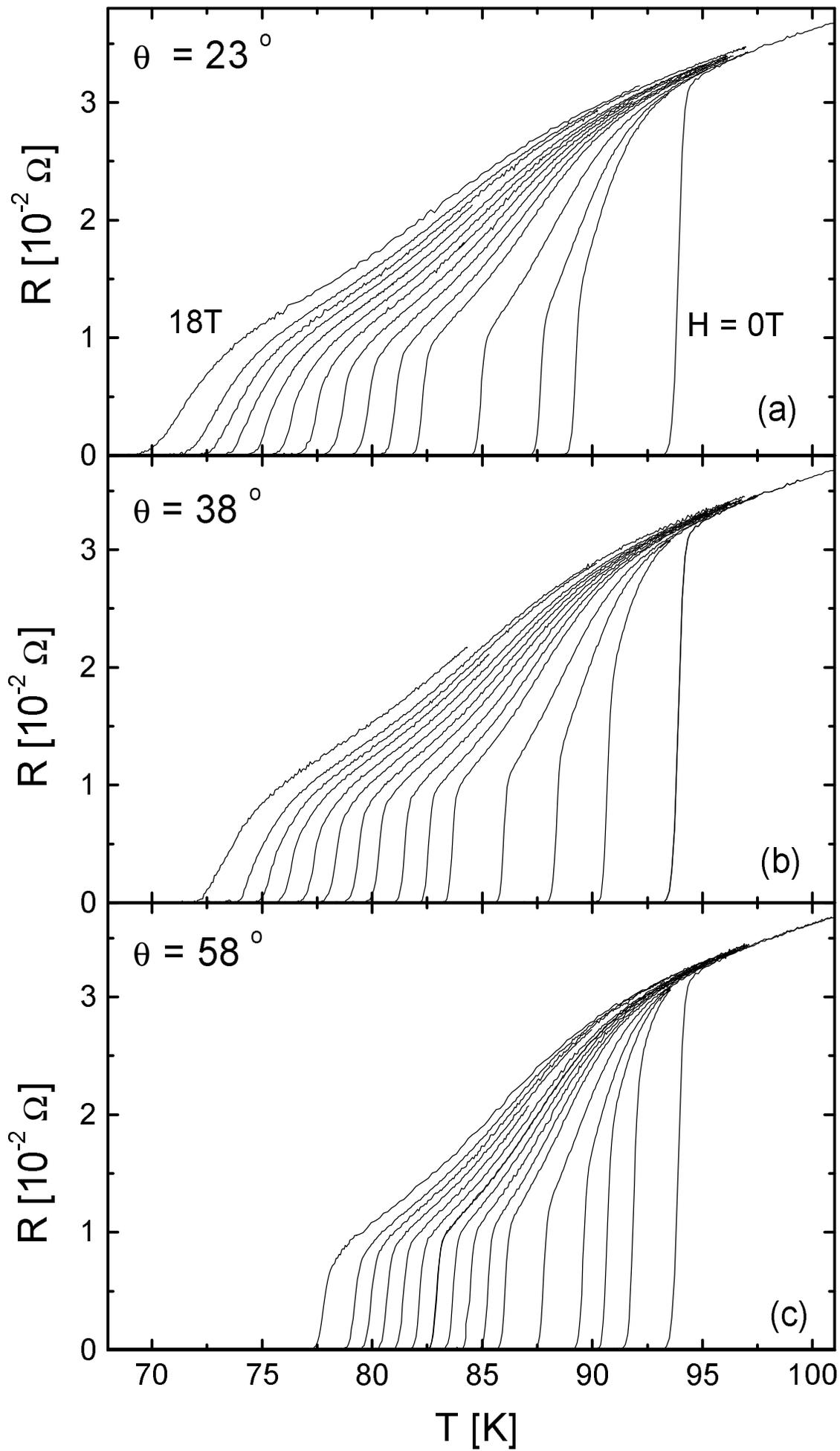
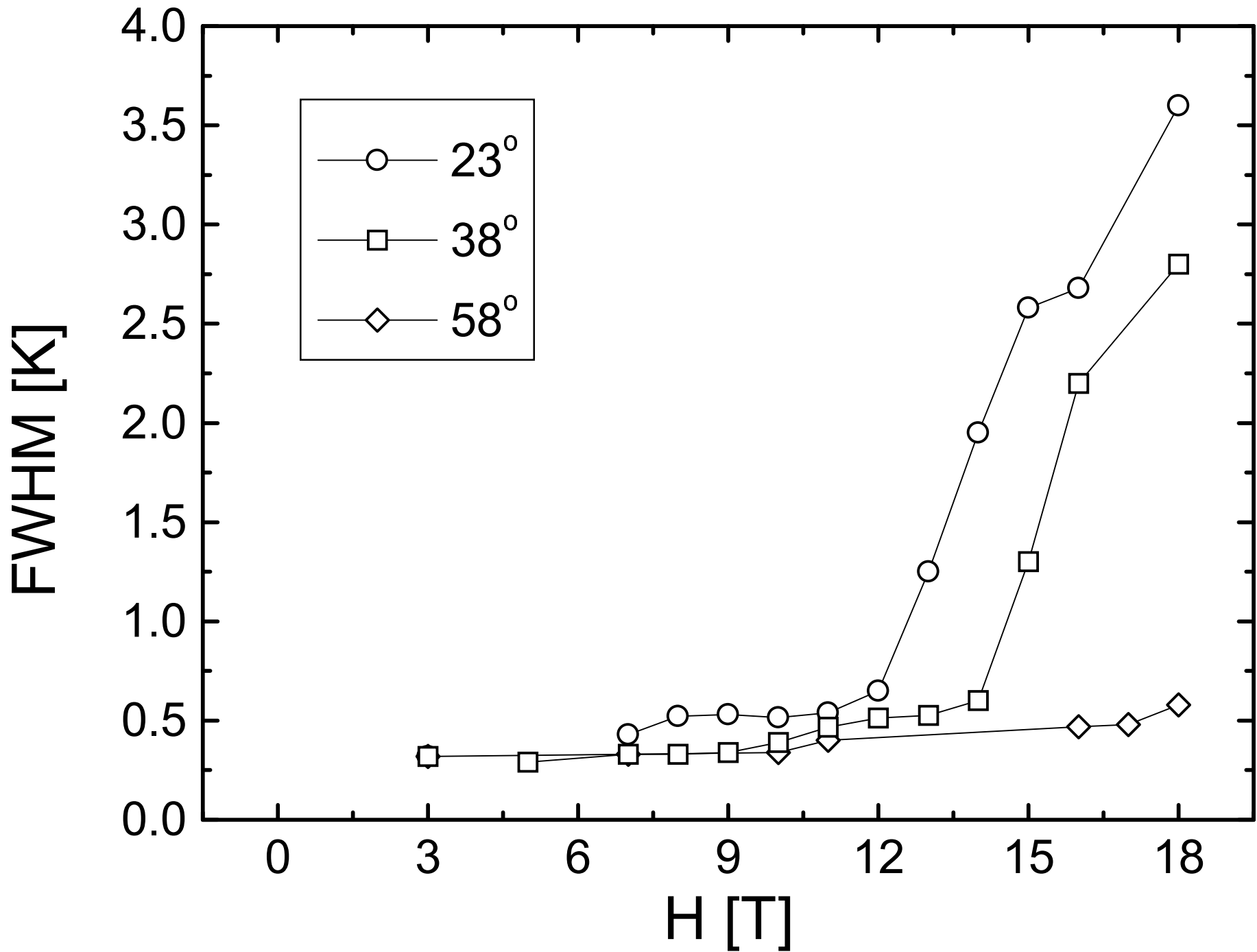
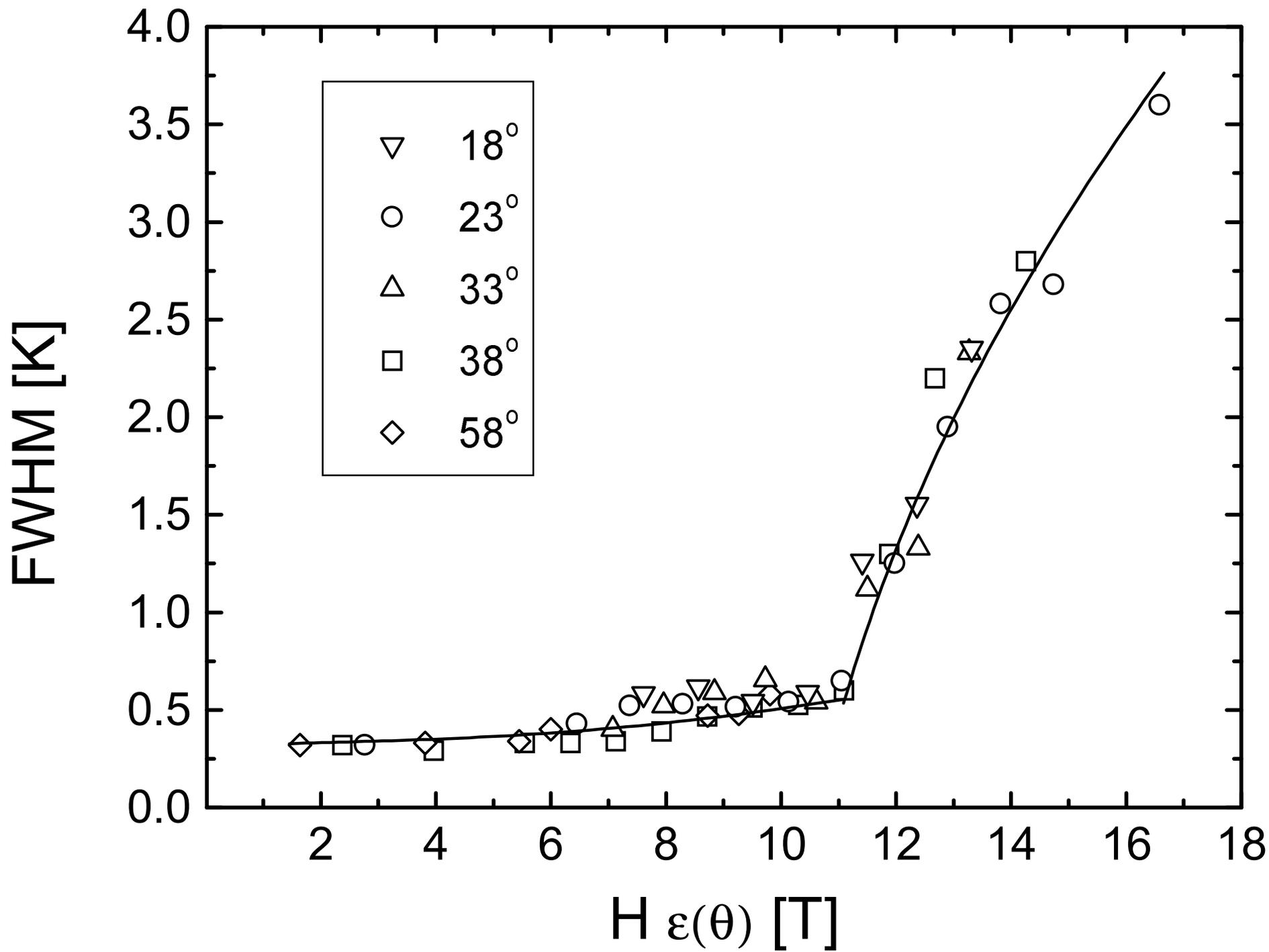
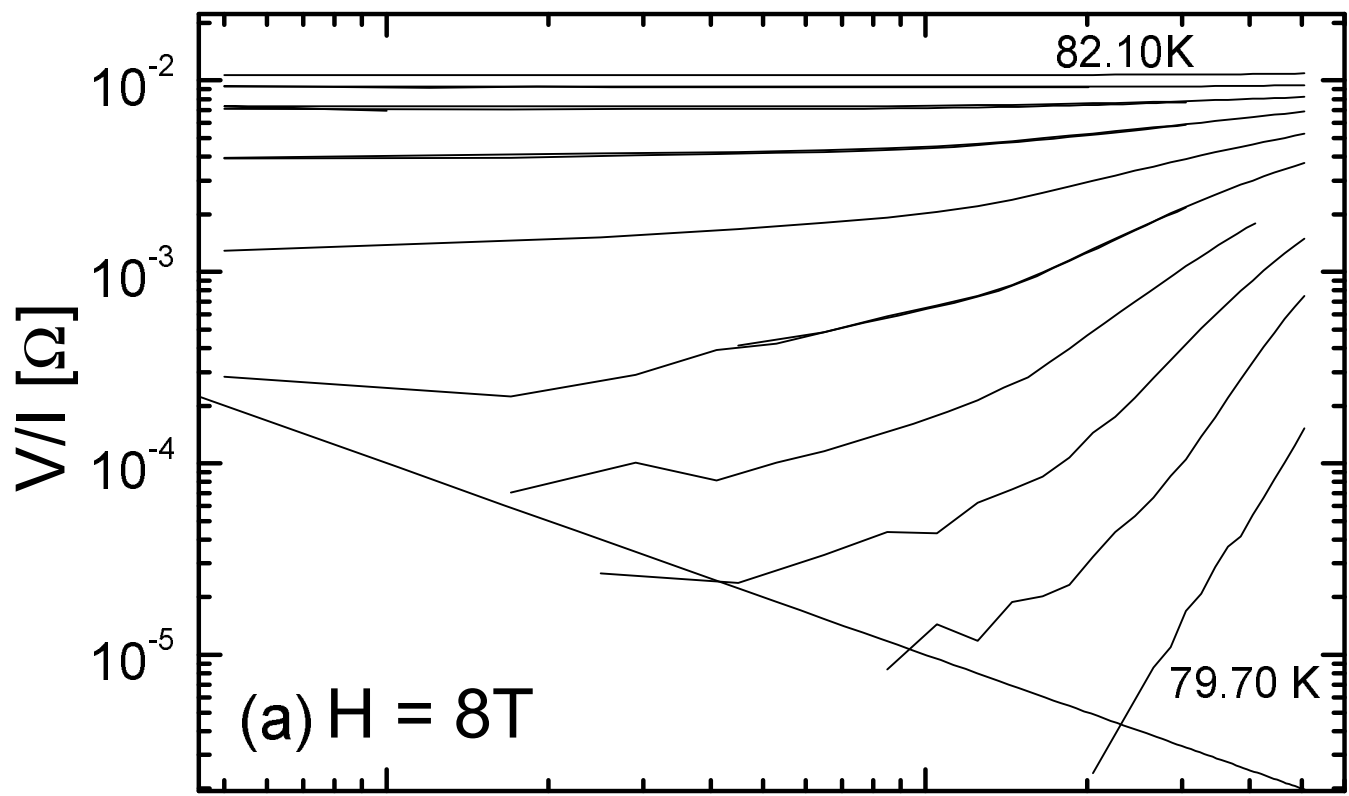
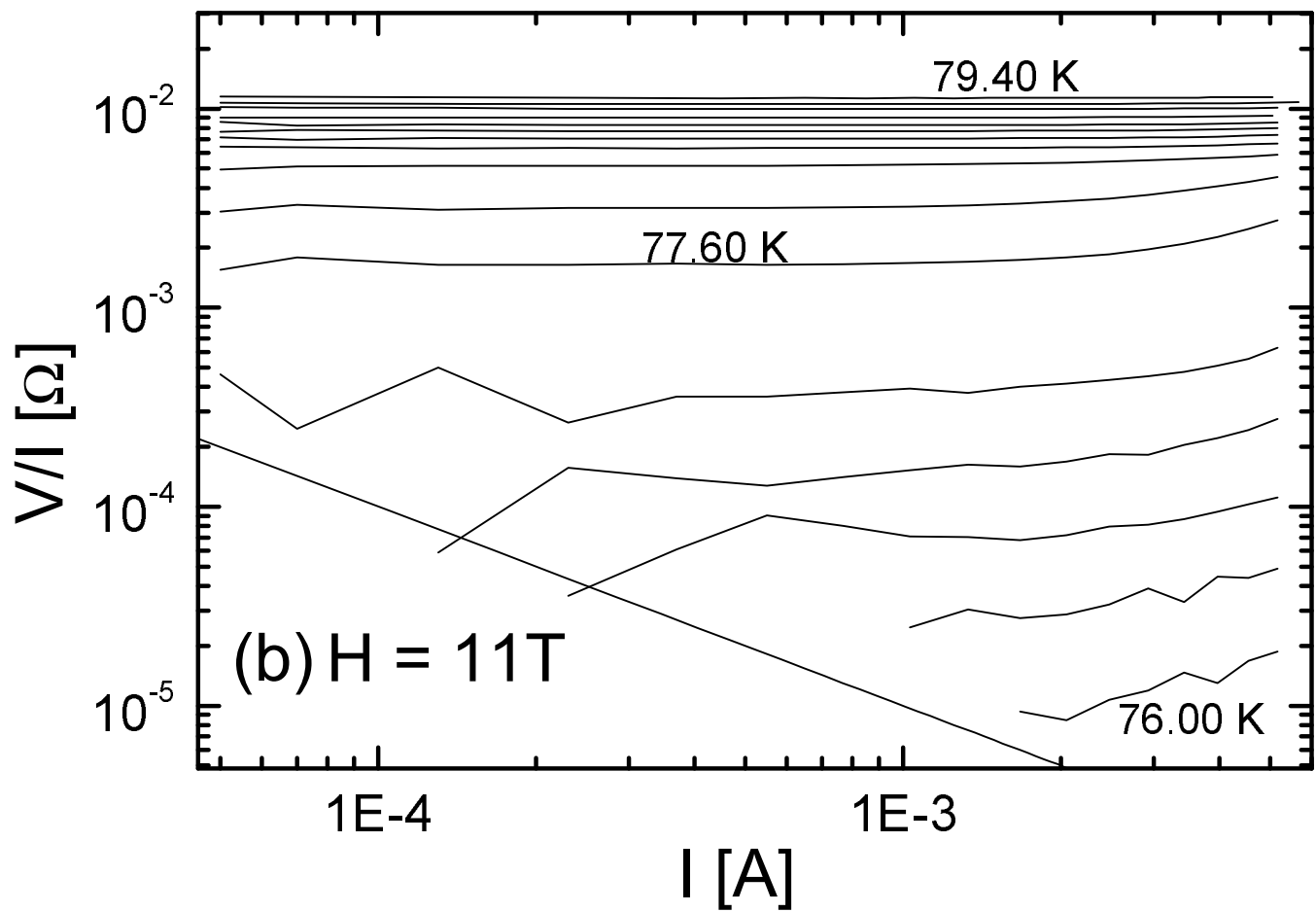
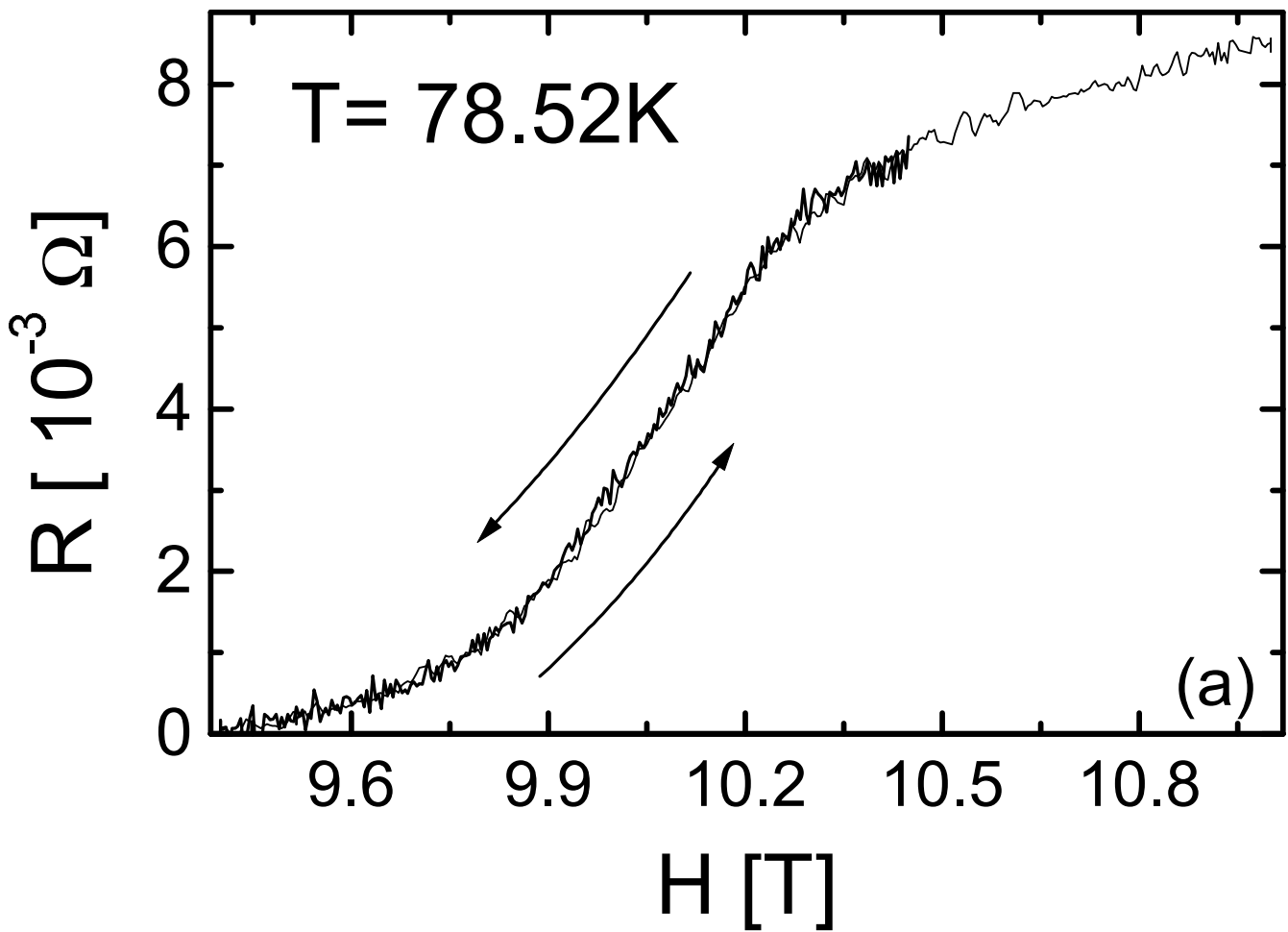
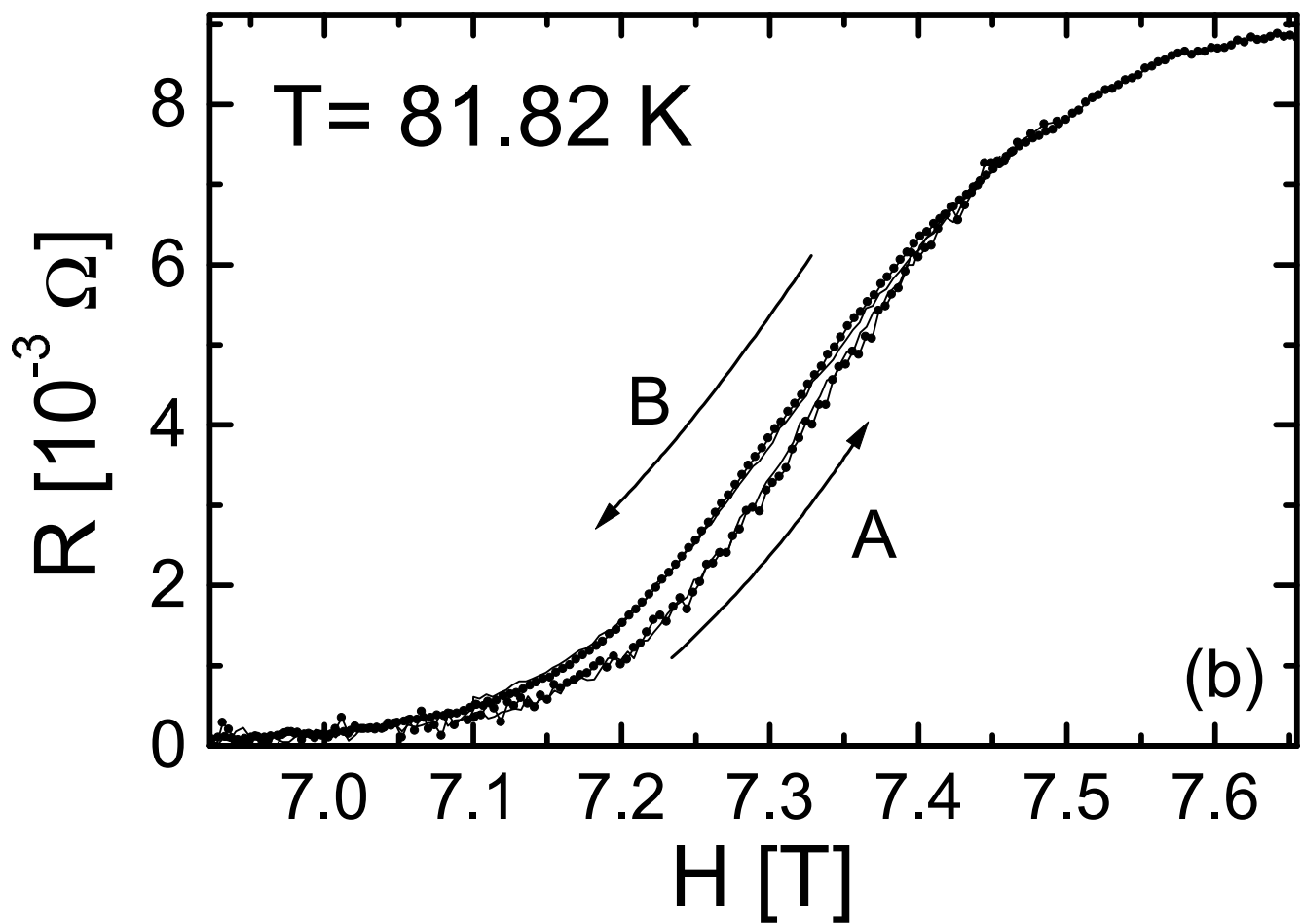
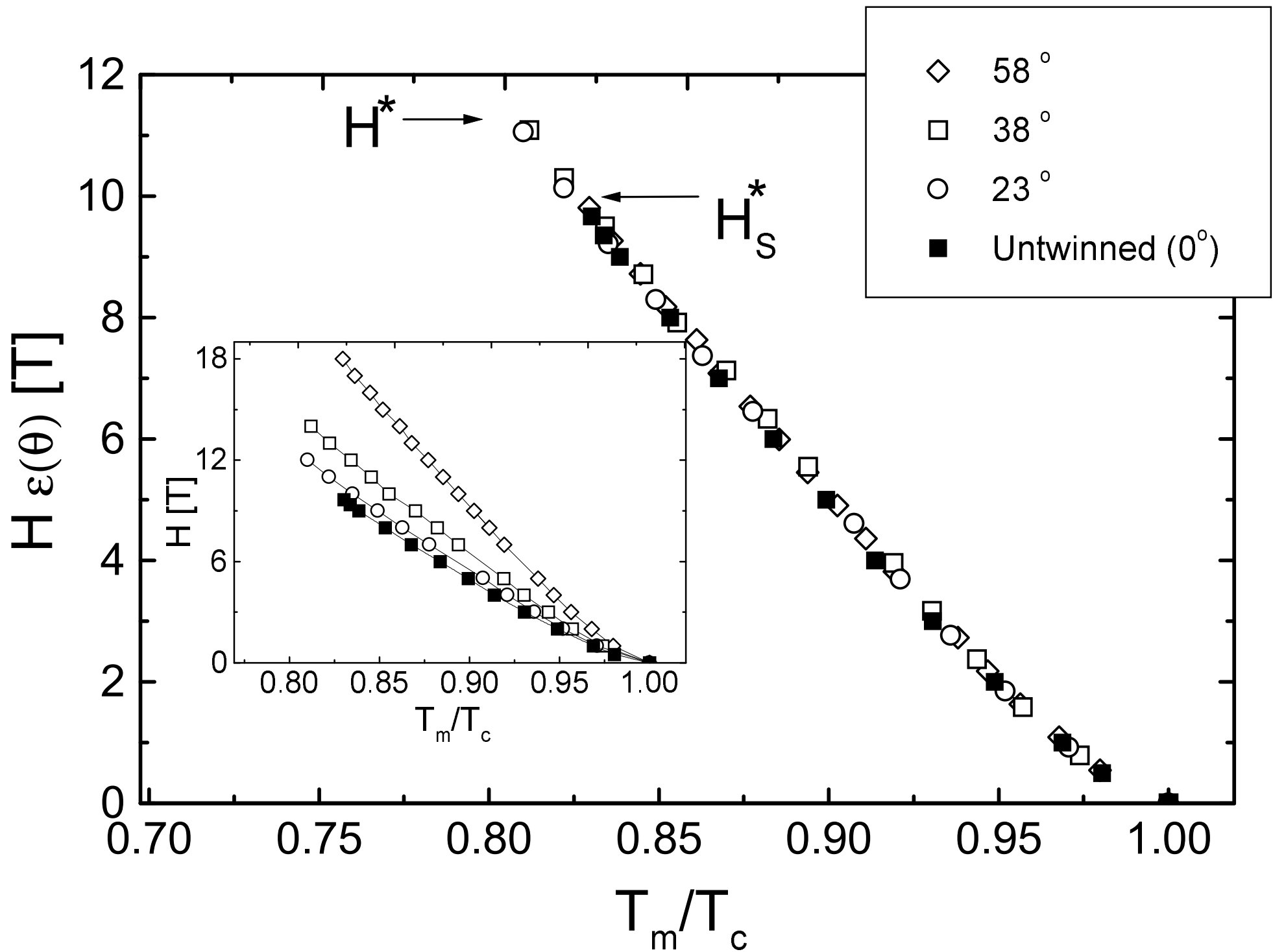
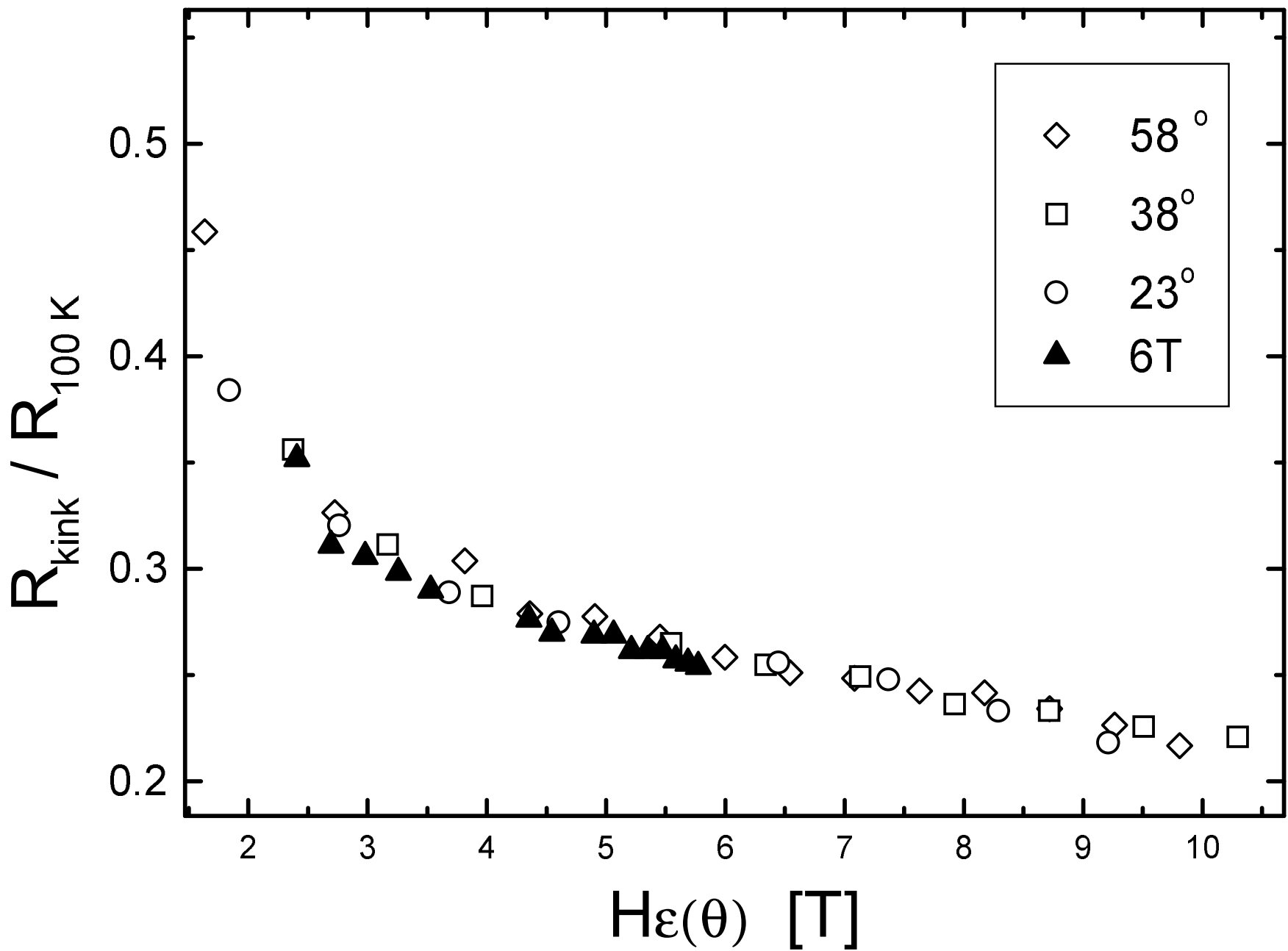
\end{document}